\documentclass[journal]{IEEEtran}
\usepackage{generic}
\usepackage{cite}
\usepackage{amsmath,amssymb,amsfonts}
\usepackage{algorithmic}
\usepackage{graphicx}
\usepackage{xcolor}
\usepackage{stfloats}
\usepackage{textcomp}
\usepackage{diagbox}
\usepackage{tabularx}
\usepackage[normalem]{ulem}
\usepackage{colortbl}
\usepackage{longtable}
\usepackage{float}
\usepackage{textcomp}

\ifCLASSINFOpdf
\else
\fi
\begin{document}
%
\title{Multi-scale reconstruction of undersampled spectral-spatial OCT data for coronary imaging using deep learning}
%
%
%

\author{Xueshen Li, Shengting Cao, Hongshan Liu, Xinwen Yao, Brigitta C. Brott, Silvio H. Litovsky, Xiaoyu Song, Yuye Ling, Yu Gan
\thanks{This work was supported in part by National Science Foundation  (CRII-1948540), New Jersey Health Foundation, the  National  Center  for  Advancing Translational  Research  of  the  National  Institutes  of  Health  underaward  number  UL1TR003096.}
\thanks{X. Li, H. Liu, and Y. Gan were with Electrical and Computer Engineering Department at the University of Alabama and are with the Biomedical Engineering Department at Stevens Institute of Technology (e-mail: ygan5@stevens.edu).}
\thanks{S. Cao is with the Electrical and Computer Engineering Department at the University of Alabama}
\thanks{X. Yao was with Institute for Health Technologies, Nanyang Technological University,Singapore.}
\thanks{B. Brott and S. Litovsky are with School of Medicine, The University of Alabama at Birmingham.}
\thanks{X. Song is with Icahn School of Medicine at Mount Sinai.}
\thanks{Y. Ling is with John Hopcroft Center for Computer Science, Shanghai Jiao Tong University.}
}

%
%

\markboth{IEEE TRANSACTIONS ON BIOMEDICAL ENGINEERING, VOL. XX, NO. XX, XXXX 2022}
{Shell \MakeLowercase{\textit{et al.}}: Bare Demo of IEEEtran.cls for IEEE Journals}
%



\maketitle

\begin{abstract}
Coronary artery disease (CAD) is a cardiovascular condition with high morbidity and mortality. Intravascular optical coherence tomography (IVOCT) has been considered as an optimal imagining system for the diagnosis and treatment of CAD. Constrained by Nyquist theorem, dense sampling in IVOCT attains high resolving power to delineate cellular structures/features. There is a trade-off between high spatial resolution and fast scanning rate for coronary imaging. In this paper, we propose a viable spectral-spatial acquisition method that down-scales the sampling process in both spectral and spatial domain while maintaining high quality in image reconstruction. The down-scaling schedule boosts data acquisition speed without any hardware modifications. Additionally, we propose a unified multi-scale reconstruction framework, namely Multiscale-Spectral-Spatial-Magnification Network (MSSMN), to resolve highly down-scaled (compressed) OCT images with flexible magnification factors. We incorporate the proposed methods into Spectral Domain OCT (SD-OCT) imaging of human coronary samples with clinical features such as stent and calcified lesions. Our experimental results demonstrate that spectral-spatial down-scaled data can be better reconstructed than data that is down-scaled solely in either spectral or spatial domain. Moreover, we observe better reconstruction performance using MSSMN than using existing reconstruction methods. Our acquisition method and multi-scale reconstruction framework, in combination, may allow faster SD-OCT inspection with high resolution during coronary intervention.
\end{abstract}

\begin{IEEEkeywords}
Optical coherence tomography; Coronary imaging; Super-resolution; Deep learning
\end{IEEEkeywords}

%
\IEEEpeerreviewmaketitle

\section{INTRODUCTION}
Coronary Artery Disease (CAD) is the narrowing of coronary arteries caused by build-up of atherosclerotic plaques. It is the most common type of heart disease, leading to 1 in 7 deaths in the U.S \cite{benjamin2019heart}. Percutaneous Coronary Intervention (PCI) is one of the most common non-surgical procedures to open clogged coronary arteries. Approximately one million PCI are performed annually \cite{benjo2015high}. To assess PCI performance, imaging tools are necessary before and after the procedure, among which Intravascular Optical Coherence Tomography (IVOCT) offers a real-time three-dimensional evaluation with highest resolution \cite{wijns2015optical, meneveau2016optical, jones2018angiography}. 

Most commercial IVOCT is based on Swept Source Optical Coherence Tomography (SS-OCT), which has a high scanning rate ($>$100 frames/s) with a spatial resolution around 10 $\mu$m \cite{tearney2008imaging}. This resolution, however, is insufficient to delineate key features such as endothelial lining, stent coverage, and cholesterol crystals, at a much finer level of 2 $\mu$m for accurate assessment of stent implantation. In contrast, Spectral Domain Optical Coherence Tomography (SD-OCT) at a shorter wavelength outperforms its swept source counterpart in resolution and contrast, and has demonstrated its capability in visualizing critical structures\cite{LinboLiu.2011}. Nonetheless, resolving at 2 $\mu$m requires a denser sampling that inevitably  leads to a lower scanning rate of about 8 frames/s \cite{LinboLiu.2011}, insufficient for performing catheter pullback to visualize the length of arteries during PCI. Currently, there is no technology available to simultaneously maintain high spatial resolution at such a 2 $\mu$m level and a fast scanning rate. A real-time visualization with higher resolution will be desirable to better PCI outcome.

Real-time visualization requires both fast acquisition and display. A cost-effective approach to improve scanning rate is to sample and store less data. Spectral domain-OCT acquires data and processes data for visualization in spectral domain. Various methods have been proposed to down-scale OCT data from either spatial or spectral domain \cite{Fang.2013,Fang.2017,Abbasi.2018,Daneshmand.2021, Zhang.2021b, KaichengLiang.2020, Z.Yuan.2020}. For example, low signal-to-noise ratio (SNR) OCT images from spatial undersampling were obtained and reconstructed via sparse representation \cite{Fang.2013,Fang.2017,Abbasi.2018,Daneshmand.2021}. Also, spectral raw data was undersampled and reconstructed using machine learning or deep learning (DL) methods\cite{Zhang.2021b,KaichengLiang.2020, Z.Yuan.2020}. However, these methods shared downsampling factors up to 4-fold, still insufficient to booster acquisition and display from the level of 8 frames/s to the level of 100 frames/s for coronary intervention \cite{tearney2008imaging}.

\begin{figure}[t]
\centering
\includegraphics[width=8 cm]{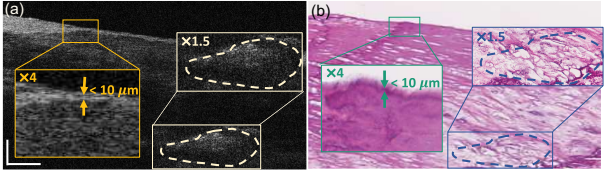}
\caption{Representative of a pairwise match between coronary OCT image (left) and its histology (right). Two regions are magnified with different scale factors for coronary analysis. scale bar: 200 $\mu$m.}
\label{fig:multi}
\vspace{-5mm}
\end{figure}

Moreover, multi-scale magnification is needed during reconstruction of down-scaled images. An example is shown in Fig. \ref{fig:multi}, where one region of fibrous cap in human coronary is shown in an OCT cross-sectional image with its matched histology. There are two regions of clinical interest: the endothelial layer (gold box in (a) and green box in (b)) and the lipid pool (white box in (a) and blue box in (b)). Based on different needs, the two regions are magnified at different scales. As gatekeepers for passage of lipoprotein and leukocytes into intima, the endothelial layer needs to be magnified with a larger scale for accurate thickness assessment. For lipid pools, however, thickness assessment is of less interest thus a lower magnification suffices. A binary justification of whether the distance between pool edge to lumen is larger than 65 $\mu$m or not suffices the evaluation of risk of rupture\cite{Tearney.2012}. 
Resolving images to the finest resolution, unfortunately, will increase the total number of pixels in display and add computational burden for visualization. As a result, a multi-scale magnification framework would be beneficial for real-time coronary visualization and analysis. At present, multi-scale magnification is illusive for OCT image reconstruction. 


Existing reconstruction methods \cite{KaichengLiang.2020, Z.Yuan.2020} treat each scale factor as a single task and only work for scale factors that are integers. Whenever a new scale factor is needed, a new model is needed and requires re-training. As a result, multiple neural networks and individual training are necessary if following existing methods to maintain the flexibility with multi-scale magnification. A unified neural network model with high flexibility in magnification factors and computational cost-effectiveness is preferable.

In this paper, we devise an efficient spectral-spatial domain acquisition method for SD-OCT systems. Our method reduces data size in both spectral and spatial domains during imaging. Additionally, we propose a unified neural network, namely Multiscale-Spectral-Spatial-Magnification Network (MSSMN) for image reconstruction. Inspired by super-resolution technology, MSSMN magnifies compressed information in both spectral and spatial domains with any arbitrary magnification factors. With extensive experiments in human coronary OCT images, we demonstrate high accuracy, data efficiency, and magnification flexibility of our proposed method. Our contributions are as below:
\begin{itemize} 
\item We propose a novel method to reduce data size by down-scaling SD-OCT data in both spectral domain and spatial domain. The proposed method improves acquisition speed. Our method does not require any additional optical setup and can be easily integrated into existing SD-OCT systems.
\item We devise a unified network, MSSMN, to address the issue of multi-scale magnification. To the best of our knowledge, it is the first multi-scale OCT reconstruction framework in spectral-spatial domain and allows for arbitrary magnification factors.
\item We propose the first super-resolution neural network for coronary OCT imaging with an aim to highlight pathological features such as calcified region, stent strut, and lipid region.
\end{itemize}

\begin{figure*}[t]
\centering
\includegraphics[width=18cm]{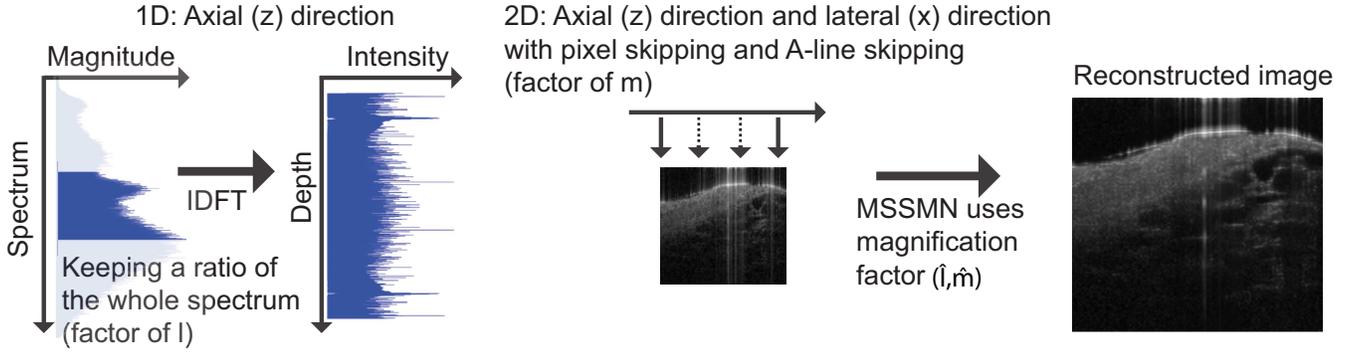}
\caption{Illustration of the proposed spectral-spatial acquisition method and multi-scale reconstruction framework. During scanning, the A-lines are skipped in lateral direction at factor of $m$. When acquiring depth information through A-lines, spectrum data is truncated at factor of $l$. After IDFT, we skip pixels in spatial domain at factor of $m$. The spectrum truncation and pixel skipping reduces data storage consumption; the A-line skipping boosts image acquisition speed. The structure of MSSMN is detailed in Fig. \ref{fig:picture003}.}
\label{fig:picture002}
\vspace{-0.5cm}
\end{figure*}

\section{RELATED WORK}

\subsection{OCT image acquisition/compression}

In SD-OCT, images can be compressed by removing part of their spectrum data\cite{AntoniaLichtenegger.2021,Zhang.2021b}. Lichtenegger et al. acquired discontinuous spectrum by modifying light source\cite{AntoniaLichtenegger.2021}; Zhang et al. studied 2-fold and 3-fold undersampled spectral data\cite{Zhang.2021b}. In spatial domain, previous research aimed to compress OCT images without modifying the spectrum. Algorithms based on transform coding, such as discrete cosine transform (DCT) and discrete wavelet transform (DWT), have been used in OCT image compression\cite{DBLP:journals/corr/SimonyanZ14a}. Also, algorithms based on sparse representation were also applied to OCT image compressing\cite{Fang.2013,Fang.2017}. The U-Net \cite{Ronneberger.2015} structure has been used in compressing OCT images because of their unique bottleneck layers\cite{PengfeiGuo.2020}. 
\subsection{OCT image reconstruction}
\subsubsection{None-DL methods for OCT image reconstruction}
Sparse representation algorithms have been applied to reconstruct OCT images\cite{Fang.2013,Fang.2017}. Recently, a nonlocal weighted sparse representation (NWSR) method was proposed\cite{Abbasi.2018}. The NWSR algorithm achieved better estimation of sparse representation by exploiting information from noisy and denoised patches. Very recently, a mixed low rank approximation and second-order tensor-based total variation (LRSOTTV) approach was presented for OCT\cite{Daneshmand.2021}. Compared with other low-rank tensor-based methods, the LRSOTTV achieved better results and lower complexity by adopting a tensor prior.

\subsubsection{DL-based methods for OCT image reconstruction}
Variations of U-Net have been adopted in reconstructing OCT images \cite{Xu.2018,Hao.2020b,KaichengLiang.2020,Qiu2020N2NSROCTSD,AntoniaLichtenegger.2021,Zhang.2021b}. Hao et al. reconstructed high quality OCT images from their low bit-depth counterparts using a U-Net based architecture\cite{Hao.2020b}. Also with the U-Net architecture, Liang et al. enhanced the resolution while recovering realistic speckles in the OCT images\cite{KaichengLiang.2020}. Further, Qiu et al. achieved simultaneously denoising and reconstruction of OCT images using a U-Net based network\cite{Qiu2020N2NSROCTSD}. The U-Net based networks have been used in reconstructing OCT images compressed in spectral domain.

Apart from the U-Net based DL models, other DL architectures such as SRResNet, residual dense network, deep back-projection network, and residual-in-residual dense network have been used in different research of reconstructing OCT images\cite{HongmingPan.2020b, S.Cao.2020,Z.Yuan.2020}. Furthermore, the Generative adversarial network (GAN) based networks were also incorporated in many OCT reconstruction research\cite{Hao.2020b, HongmingPan.2020b,S.Cao.2020,Z.Yuan.2020,Zhou.2020,Das.2020}. Among these research, Yuan et al. proposed to super-resolve optical resolution of OCT images using DL algorithms\cite{Z.Yuan.2020}; Cao et al. increased both optical and digital resolutions of OCT\cite{S.Cao.2020}. However, none of previous research managed to develop a unified framework that was capable of reconstructing OCT images in both spectral and spatial domains by multiple magnification factors. 

\section{METHOD}

In this part, we first discuss spectral-spatial OCT image acquisition in detail. Then we define the problem of multi-scale OCT reconstruction (magnification) in spectral-spatial domain. Finally, we introduce the framework and implementations. 
\subsection{OCT acquisition in spectral-spatial domain}

We collaboratively reduce the amount of data in both spatial domain and spectral domain during OCT image acquisition with a viable scale factor of ($l$, $m$), where $l$ is a acquisition factor in spectral domain and $m$ is a acquisition factor in spatial domain. As shown in Fig. \ref{fig:picture002}, in spectral domain, we only collect $\frac{1}{l}$ of the spectrum at the center of the bandwidth, since it carries most information spectrum. In spatial domain, we skip $m-1$ A-lines per $m$ A-lines during scanning. The A-line skipping is carried out in lateral direction; in order to further reduce data storage consumption and maintain the aspect ratio of compressed OCT images in spatial domain, we keep $\frac{1}{m}$ pixels in axial direction after Inverse Discrete Fourier Transform (IDFT). 
The proposed method can be integrated with existing workflow of SD-OCT without any modifications of hardware. 

\begin{figure*}[h]
\centering
\includegraphics[width=18cm]{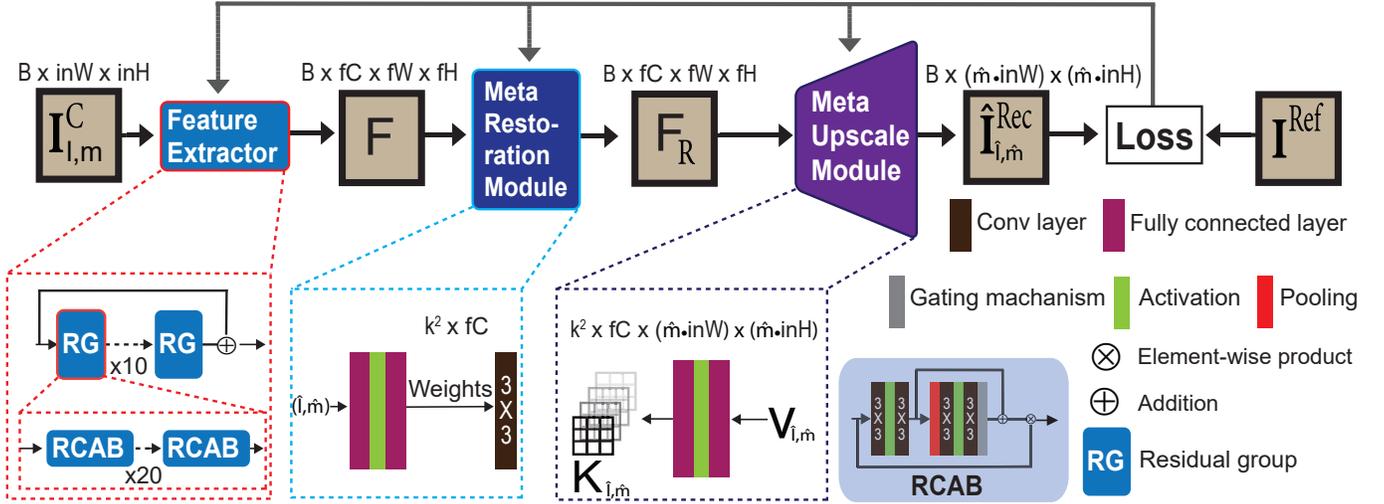}
\caption{Scheme of the proposed Multiscale-Spectral-Spatial-Magnification Network (MSSMN) using Residual Channel Attention Network (RCAN) as feature extractor. Residual Channel Attention Blocks (RCAB) are the building block of RCAN. The key components of our framework are feature extractor, meta-upscale module, and meta-restoration module. \textcolor{black}{The feature map $F$ is extracted by feature extractor, then processed by meta-restoration (generating $F_{R}$) and meta-upscaling modules (generating $\hat{I}^{Rec}_{\hat{l},\hat{m}}$). Magnification factors $(\hat{l},\hat{m})$ determine the reconstruction weights. The width and height of reconstructed OCT images are magnified by spatial magnification factor $\hat{m}$. Number of feature channels fC $=64$ and convolution kernel size $k=3$.}}
\label{fig:picture003}
\end{figure*}
\textcolor{black}{With less data sampled, such acquisition process could be considered as a degradation process. The data-acquisition and reconstruction process are degradation and super-resolution processes \cite{W.Yang.2019}. In SD-OCT images, the degradation process can be reformulated as: }
\begin{equation}
I^{C}_{l,m} = \mathcal{F}^{-1}(S \cdot T_{l})\downarrow_{m}
\end{equation}
where $I^{C}_{l,m}$ is a OCT image compressed with a acquisition factor $(l,m)$ in spectral-spatial domain; $\mathcal{F}^{-1}$ stands for inverse Fourier transform operator; $S$ is the raw spectrum acquired by SD-OCT systems; $T_{l}$ is a truncating kernel which keeps $\frac{1}{l}$ of the spectrum in spectral domain; $\downarrow_{m}$ stands for skipping operation in both axial and lateral directions in spatial domain, keeping $\frac{1}{m}$ pixels in each direction. Note that the truncating operation in the spectral domain is equivalent to the convolution in the spatial domain because of the convolution theorem. $T_{l}$ and $\downarrow_{m}$ are simple operations with time complexity of $O(1)$. We refer to OCT images acquired by $(l,m)$ with $l>1$ and $m>1$ as low resolution (LR) image.

\subsection{Multi-scale magnification of compressed OCT images}
The acquisition process of compressed OCT images is deterministic; once the acquisition factors $l$ and $m$ are fixed, a unique $I^{C}_{l,m}$ is generated. Reconstruction of OCT images can be considered as a reversing process of compression, which provides high quality images given $I^{C}_{l,m}$. Ideally, the output after reconstruction should be $I^{Ref}=\mathcal{F}^{-1}(S)$. However, reconstruction is a ill-posed inverse problem since multiple solutions may exist for a single $I^{C}_{l,m}$. The DL-based algorithm learns a single non-linear mapping from $I^{C}_{l,m}$ to ${\hat{I}}^{Rec}$.
In this paper, we rather propose a novel reconstruction framework for resolving compressed OCT images by multiple magnification factors:
\begin{equation}
{\hat{I}}^{Rec}_{\hat{l},\hat{m}} = G_{\hat{l},\hat{m}}(I^{C}_{l,m}; \theta_{\hat{l},\hat{m}})
\label{3}
\end{equation}
here $G_{\hat{l},\hat{m}}$ is a unified neural network which has input of LR image and two magnification factors: magnification factor $\hat{l}$ in spectral domain and magnification factor $\hat{m}$ in spatial domains; $\theta_{\hat{l},\hat{m}}$ stands for parameters in $G_{\hat{l},\hat{m}}$. 


\subsection{Unified deep learning framework for OCT image magnification}
Our framework of MSSMN consists of the following modules: feature extractor, meta-upscaling and meta-restoration modules\cite{Xu.2018}, and loss function.

\subsubsection{Feature extractor}
\textcolor{black}{The feature map $F$ is extracted from $I^{C}_{l,m}$, using feature extractor $E$ which contains parameters $\theta_{E}$.}
We choose to use Residual Channel Attention Network (RCAN) for feature extraction, which adopts the Residual-In-Residual (RIR) structure \cite{YulunZhang.2018b}. The RCAN achieves much deeper network structure\cite{Wang.2019b}.




\subsubsection{Meta-upscaling and meta-restoration modules for multiple degradation parameters}
The ${\hat{I}}^{Rec}_{\hat{l},\hat{m}}$ is reconstructed from the feature map $F$. We develop a unified framework that resolves compressed OCT images in multiple magnification factors. For this purpose, we adopt the meta-upscaling module\cite{Hu2019MetaSRAM} and meta-restoration module\cite{X.Hu.2020}. 
\textcolor{black}{The magnification factors ($\hat{l},\hat{m}$) are used to determine the weights of the convolution kernels in meta-learning modules. The meta-restoration module $\varphi_{R}$ dynamically predicts weights $W_{R}$ of a regular convolution layer based on magnification factors $(\hat{l},\hat{m})$. $W_{R}$ possesses a dimension $k^{2}\times$ fC. The convolution layer restores $F$ and generates restored feature map $F_{R}$. Two feature maps, $F$ and $F_{R}$, have the same dimension of B$\times$fC$\times$fW$\times$fH, where B is the number of patches. $\theta_{\varphi_{R}}$ denotes parameters in $\varphi_{R}$.}
The meta-upscaling module $\varphi_{U}$ dynamically predicts weights of kernels $K_{\hat{l},\hat{m}}$, given magnification factors $(\hat{l},\hat{m})$. \textcolor{black}{$K_{\hat{l},\hat{m}}$ has a shape of $k^{2}\times$fC$\times (\hat{m}\cdot$inW$)\times(\hat{m}\cdot$inH$)$}. With $F_{R}$ and $K_{\hat{l},\hat{m}}$, the ${\hat{I}}^{Rec}_{\hat{l},\hat{m}}$ is reconstructed by a feature mapping function $\Phi$:
\begin{equation}
\begin{split}
    K_{\hat{l},\hat{m}}(w,h) & =\varphi_{U}(v_{\hat{l},\hat{m}};\theta_{\varphi_{U}})
    \\
    {\hat{I}}^{Rec}_{\hat{l},\hat{m}}(w,h) & = \Phi(F_{R}(w',h'), K_{\hat{l},\hat{m}}(w,h))
    \\
    (w',h') &= (\lfloor \frac{w}{\hat{m}} \rfloor,\lfloor \frac{h}{\hat{m}} \rfloor)
\end{split}
\end{equation}
where $\theta_{\varphi_{U}}$ denotes parameters in $\varphi_{U}$, $(w, h)$ denotes locations of pixels in $I^{C}_{l,m}$, $(w',h')$ denotes locations of pixels in $F_{R}$, $v_{\hat{l},\hat{m}}$ is the input vector relating to magnification factors $(\hat{l},\hat{m})$ and size of $I^{C}_{l,m}$, $\lfloor \cdot \rfloor$ stands for floor operator. The $K_{\hat{l},\hat{m}}(w,h)$ is a $k \times k$ kernel with its center at $(w',h')$. Note that for a single $(w',h')$ in $F_{R}$, there will be multiple kernels $K_{\hat{l},\hat{m}}(w,h)$ for reconstructing the feature map with varying magnification factors $(\hat{l},\hat{m})$.
\textcolor{black}{We set number of feature channels fC $=64$ and convolution kernel size $k=3$.}

The $v_{\hat{l},\hat{m}}$ and $\Phi$ are defined as:
\begin{equation}
\begin{split}
v_{\hat{l},\hat{m}}(w,h) = (\frac{w}{\hat{m}}-\lfloor \frac{w}{\hat{m}}\rfloor, \frac{h}{\hat{m}}-\lfloor \frac{h}{\hat{m}}\rfloor, \hat{l})
\\
\Phi(F_{R}(w',h'),K_{\hat{l},\hat{m}}(w,h)) = F_{R}(w',h') \otimes K_{\hat{l},\hat{m}}(w,h)
\end{split}
\end{equation}

Based on this meta-learning scheme, we magnify compressed OCT images by multiple scale factors with a unified network.

\subsubsection{Loss function}
\textcolor{black}{Mean squared error (MSE) is widely used loss function for general
image restoration. However, the L1 (Mean absolute error) has demonstrated potential of achieving better convergence \cite{7797130}. We adopted the L1 loss for training all models. }

\section{EXPERIMENTS AND RESULTS}

\subsection{Experimental dataset}
Human coronary samples were collected from the School of Medicine at the University of Alabama at Birmingham (UAB) and delivered to the University of Alabama for imaging. The autopsy specimens were de-identified and not considered as human subjects, in compliance with UAB’s Institutional Review Board (IRB). Specimens were imaged via a commercial OCT system (Thorlabs Ganymede, Newton, NJ) \cite{S.Cao.2020}. We collected $2996$ OCT images (B-scans) from 23 specimens. The depth of OCT images were $2.56$ mm. The width of the images ranged from $2$ mm to $4$ mm depending on actual sample size. Among all images, the pixel size remained $2$ $\mu$m $\times$ $2$ $\mu$m within a B-scan. Samples were processed for Hematoxylin-Eosin (H\&E) histology after imaging. 

We acquired the OCT images using the proposed spectral-spatial acquisition method. In spectral domain, we used $\frac{1}{2}$, $\frac{1}{3}$ and $\frac{1}{4}$ of spectrum data (denoted as $l=2,3,4$ respectively). In the spatial domain, we used $m=2,3,4$ for A-line skipping during scanning and pixel skipping after IDFT. The experimental dataset was divided into $5$ folds by ID of specimen for cross-validation. 

\begin{figure*}[h]
\centering
\includegraphics[width=18cm]{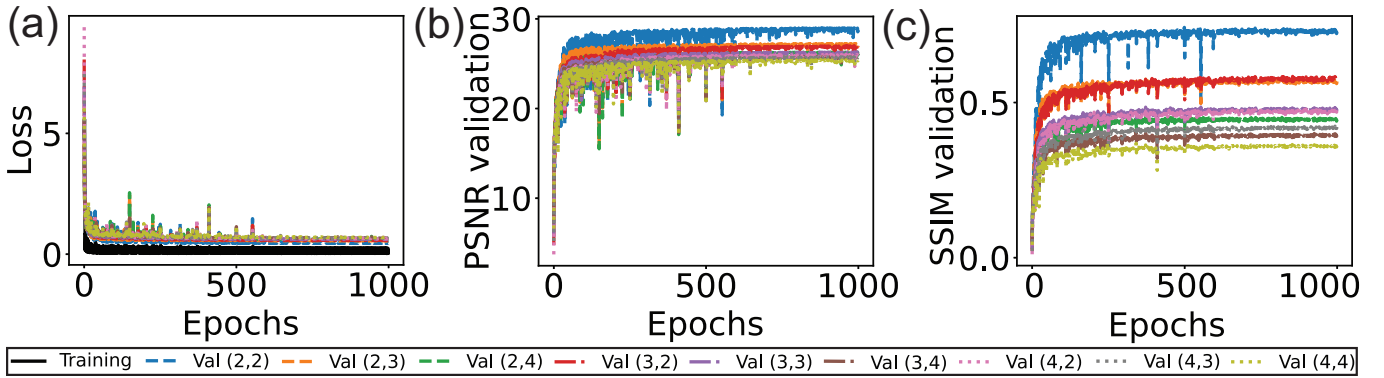}
\caption{Representative loss function plots and validation PSNR and SSIM scores. (a) Training and validation loss values over epochs. (b) PSNR scores over epochs in validation dataset. (c) SSIM scores over epochs in validation dataset. }
\label{fig:lossFunction}

\end{figure*}

\subsection{Network implementation}
\textcolor{black}{We use RCAN as the feature extractor to implement MSSMN}. As a comparative study, we implemented reconstruction networks with other feature extractors, including Multi-Scale Residual Network (MSRN)\cite{Li.2018c}, Residual Dense Network (RDN)\cite{Zhang2018ResidualDN}, and Residual Dense Unet (RDU)\cite{GurrolaRamos.2021}. For the network implementations, we used 8 multi-scale residual blocks for MSRN; 10 residual dense blocks (RDB) for RDN, with each RDB containing 6 convolution layers; 7 levels for RDU, with each level containing 1 denoising block; 10 residual groups (RG) for RCAN, with each RG containing 20 residual channel attention blocks (RCAB). We implemented the sub-pixel upsampling module \cite{Shi2016RealTimeSI} for each features extractors. 
\textcolor{black}{In comparison with the abovementioned network structure, RCAN achieves deep trainable network structure and is capable of adaptively learn useful channel-wise features \cite{YulunZhang.2018b}}.

\subsection{Training details}
The intensities were scaled to $[0,1]$. For each network, the training procedure was repeated $5$ times on different data folds. 
\textcolor{black}{During the training, we randomly extract 16 non-overlapped patches with the size of $64\times64$. Those patches are the input of the model. We augment the patches by randomly flipping to prevent overfitting.} Within each batch, magnification factors $(\hat{l},\hat{m})$ for MSSMN remained the same. We used Adam optimizer\cite{kingma2014adam} for training. The learning rate was initialized as $10^{-4}$, following by a half decay for every $200$ epochs. \textcolor{black}{In total the networks were trained 1000 epochs to ensure convergence.} \textcolor{black}{The experiments were carried out in parallel on three RTX A6000 GPUs.} \textcolor{black}{Training on each data fold take around 10 hours.}

\begin{table}[t]
\centering
\caption{Results Of Reconstruction from OCT Image of acquisition factor $(l,m)$ given a compression rate of 25\%. All Results Are Averaged Based On Five-Fold Cross-Validation. \textcolor{red}{Red} Indicates The Best Performance.}
\label{table:tabSSC}
\begin{tabular}{ll|l|c} 
\hline
\diagbox{Metrics}{$(l,m)$}                           & \multicolumn{1}{c}{(1,4)}                               & \multicolumn{1}{c}{(4,1)}                               & (2,2)                                                                                                          \\ 
\hline
\begin{tabular}[c]{@{}l@{}}PSNR\\SSIM\end{tabular} & \begin{tabular}[c]{@{}l@{}}25.8183\\0.4860\end{tabular} & \begin{tabular}[c]{@{}l@{}}26.6053\\0.4622\end{tabular} & \multicolumn{1}{l}{\begin{tabular}[c]{@{}l@{}}\textcolor{red}{29.1390}\\\textcolor{red}{0.7570}\end{tabular}}  \\
\hline
\end{tabular}

\end{table}

\subsection{Evaluation metrics}
We used peak signal-to-noise ratio (PSNR) and structural similarity (SSIM) \cite{ZhouWang.2004b} to measure the quality of image magnification. 
The PSNR calculates pixel-wise differences between the reconstructed image and the reference image; the SSIM focuses on structural similarly between reconstructed image and the reference image.

\subsection{Network convergence}

\textcolor{black}{For each fold, we trained the MSSMN for 1000 epochs to ensure convergence. Fig. \ref{fig:lossFunction} shows representative plots of training loss and validation loss,  PSNR and SSIM scores, all from one fold of validation dataset. The MSSMN is capable of multi-scale reconstruction. We calculated validation loss values and PSNR and SSIM scores for each magnification factor. Within 1000 epochs, the MSSMN converges for each magnification factor.}

\subsection{The efficiency of spectral-spatial compression}

We fixed a compression ratio of 25$\%$ and compared the reconstruction performance from OCT images among three schemes: i) solely in spectral domain $(l=4,m=1)$; ii) solely in spatial domain $(l=1,m=4)$; iii) jointly in spectral and spatial domains ($l=2,m=2$). In Table \ref{table:tabSSC}, we show the PNSR and SSIM of magnifying OCT images of acquisition factors $(1,4)$, $(4,1)$ and $(2,2)$ using MSSMN. 
The joint acquisition, with a factor of $(2,2)$, provides much higher scores after reconstruction, comparing to that of $(1,4)$ and $(4,1)$. Thus, acquiring OCT images in the spectral-spatial domain is a better strategy than pushing the compression limit in a single domain.

\begin{figure}[t]
\centering
\includegraphics[width=9cm]{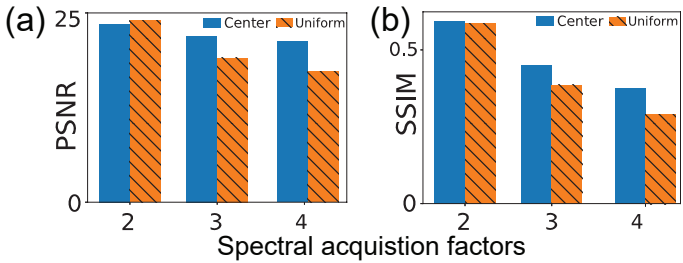}
\caption{Results of acquiring OCT images using different strategies: truncating the spectrum from center and dropping spectral data uniformly. (a): PSNR. (b): SSIM.}
\label{fig:bar}

\end{figure}

\textcolor{black}{When acquiring spectrum data, there are two possible approaches: truncating the spectrum from center and dropping spectral data uniformly. We compared the two strategies by fixing the spatial acquisition factor $m$ to 1 and varying spectral acquisition factor $l$. For the uniform acquisition, we drop every $l-1$ points in the A-line spectrum data to achieve an acquisition factor of $l$ as the same sampling strategy in \cite{Zhang.2021b}. We calculated PSNR and SSIM scores for OCT images within the whole dataset acquired from both strategies. As shown in Fig. \ref{fig:bar}, truncating the spectrum from center provides higher PSNR and SSIM scores when $l=3$ and $l=4$.}

\subsection{Performance comparison among feature extractors and meta-learning}
\begin{table*}[h]
\caption{Results Of Reconstruction From OCT Images of Acquisition Factor $(l,m)$ Using Various Feature Extractors (MSRN, RDN, RDU, RCAN) and Meta Learning (MSSMN with RCAN). All Results Are Averaged Based On Five-Fold Cross-Validation. Magnification factors $(\hat{l}, \hat{m})$ are the same as acquistion factors $(l, m)$. \textcolor{red}{Red} and \textcolor{blue}{\uline{Blue}} Indicate The Best And The Second Best Performance, Respectively.}
\label{table:tab1}
\centering
\setlength\tabcolsep{5pt}
\begin{tabular}{lll|l|l|l|l|l|l|l|l}
\hline
\diagbox{Method}{$(l,m)$} & \multicolumn{1}{l}{Metrics} & \multicolumn{1}{c}{(2,2)} & \multicolumn{1}{c}{(2,3)} & \multicolumn{1}{c}{(2,4)} & \multicolumn{1}{c}{(3,2)} & \multicolumn{1}{c}{(3,3)} & \multicolumn{1}{c}{(3,4)} & \multicolumn{1}{c}{(4,2)} & \multicolumn{1}{c}{(4,3)} & \multicolumn{1}{c}{(4,4)} \\
\hline
MSRN  & \begin{tabular}[c]{@{}l@{}}PSNR\\SSIM\end{tabular} & \begin{tabular}[c]{@{}l@{}}29.1191\\0.7566\end{tabular}                                                      & \begin{tabular}[c]{@{}l@{}}27.3316\\0.6018\end{tabular}                                                      & \begin{tabular}[c]{@{}l@{}}26.3894\\0.4866\end{tabular}                                                      & \begin{tabular}[c]{@{}l@{}}26.8260\\0.5996\end{tabular}                                                      & \begin{tabular}[c]{@{}l@{}}26.2016\\0.5061\end{tabular}                                                      & \begin{tabular}[c]{@{}l@{}}25.7575\\0.4288\end{tabular}                                    & \begin{tabular}[c]{@{}l@{}}25.8892\\0.4964\end{tabular}                                                      & \begin{tabular}[c]{@{}l@{}}25.6713\\0.4456\end{tabular}                                                      & \begin{tabular}[c]{@{}l@{}}25.3854\\0.3941\end{tabular}                                                       \\
RDN   & \begin{tabular}[c]{@{}l@{}}PSNR\\SSIM\end{tabular} & \begin{tabular}[c]{@{}l@{}}29.1173\\0.7569\end{tabular}                                                      & \begin{tabular}[c]{@{}l@{}}27.3165\\0.6023\end{tabular}                                                      & \begin{tabular}[c]{@{}l@{}}26.4040\\0.4874\end{tabular}                                                      & \begin{tabular}[c]{@{}l@{}}26.8226\\0.5967\end{tabular}                                                      & \begin{tabular}[c]{@{}l@{}}26.2199\\0.5046\end{tabular}                                                      & \begin{tabular}[c]{@{}l@{}}25.7389\\0.4302\end{tabular}                                    & \begin{tabular}[c]{@{}l@{}}25.9015\\0.4952\end{tabular}                                                      & \begin{tabular}[c]{@{}l@{}}25.6276\\0.4477\end{tabular}                                                      & \begin{tabular}[c]{@{}l@{}}25.3776\\0.3941\end{tabular}                                                       \\
RDU   & \begin{tabular}[c]{@{}l@{}}PSNR\\SSIM\end{tabular} & \begin{tabular}[c]{@{}l@{}}28.9336\\0.7526\end{tabular}                                                      & \begin{tabular}[c]{@{}l@{}}27.3247\\0.5999\end{tabular}                                                      & \begin{tabular}[c]{@{}l@{}}26.4121\\0.4870\end{tabular}                                                      & \begin{tabular}[c]{@{}l@{}}26.3479\\0.5796\end{tabular}                                                      & \begin{tabular}[c]{@{}l@{}}26.2179\\0.5065\end{tabular}                                                      & \begin{tabular}[c]{@{}l@{}}\textcolor{blue}{\uline{25.7630}}\\0.4295\end{tabular}          & \begin{tabular}[c]{@{}l@{}}25.6872\\0.4827\end{tabular}                                                      & \begin{tabular}[c]{@{}l@{}}25.5999\\0.4443\end{tabular}                                                      & \begin{tabular}[c]{@{}l@{}}25.4010\\0.3932\end{tabular}                                                       \\
RCAN  & \begin{tabular}[c]{@{}l@{}}PSNR\\SSIM\end{tabular} & \begin{tabular}[c]{@{}l@{}}\textcolor{blue}{\uline{29.1395}}\\\textcolor{blue}{\uline{0.7570 }}\end{tabular} & \begin{tabular}[c]{@{}l@{}}\textcolor{blue}{\uline{27.3534}}\\\textcolor{blue}{\uline{0.6024 }}\end{tabular} & \begin{tabular}[c]{@{}l@{}}\textcolor{blue}{\uline{26.4421}}\\\textcolor{blue}{\uline{0.4876 }}\end{tabular} & \begin{tabular}[c]{@{}l@{}}\textcolor{blue}{\uline{26.9176}}\\\textcolor{blue}{\uline{0.6023 }}\end{tabular} & \begin{tabular}[c]{@{}l@{}}\textcolor{blue}{\uline{26.2668}}\\\textcolor{blue}{\uline{0.5082 }}\end{tabular} & \begin{tabular}[c]{@{}l@{}}25.7503\\\textcolor{blue}{\uline{0.4326 }}\end{tabular}         & \begin{tabular}[c]{@{}l@{}}\textcolor{blue}{\uline{25.9547}}\\\textcolor{blue}{\uline{0.4973 }}\end{tabular} & \begin{tabular}[c]{@{}l@{}}\textcolor{blue}{\uline{25.7287}}\\\textcolor{blue}{\uline{0.4486 }}\end{tabular} & \begin{tabular}[c]{@{}l@{}}\textcolor{blue}{\uline{25.4509}}\\\textcolor{blue}{\uline{0.3947 }}\end{tabular}  \\
MSSMN & \begin{tabular}[c]{@{}l@{}}PSNR\\SSIM\end{tabular} & \begin{tabular}[c]{@{}l@{}}\textcolor{red}{29.1789}\\\textcolor{red}{0.7578 }\end{tabular}           & \begin{tabular}[c]{@{}l@{}}\textcolor{red}{27.3914}\\\textcolor{red}{0.6040 }\end{tabular}                   & \begin{tabular}[c]{@{}l@{}}\textcolor{red}{26.4562}\\\textcolor{red}{0.4876 }\end{tabular}                   & \begin{tabular}[c]{@{}l@{}}\textcolor{red}{27.0109}\\\textcolor{red}{0.6031 }\end{tabular}                   & \begin{tabular}[c]{@{}l@{}}\textcolor{red}{26.3555}\\\textcolor{red}{0.5104 }\end{tabular}                   & \begin{tabular}[c]{@{}l@{}}\textcolor{red}{25.8655}\\\textcolor{red}{0.4333 }\end{tabular} & \begin{tabular}[c]{@{}l@{}}\textcolor{red}{26.0577}\\\textcolor{red}{0.4993 }\end{tabular}                   & \begin{tabular}[c]{@{}l@{}}\textcolor{red}{25.7946}\\\textcolor{red}{0.4507 }\end{tabular}                   & \begin{tabular}[c]{@{}l@{}}\textcolor{red}{25.4995}\\\textcolor{red}{0.3978 }\end{tabular}                    \\
\hline
\end{tabular}
\end{table*}

We compare the performance of RCAN as a feature extractor with that of MSRN, RDN and RDU, as shown in Table \ref{table:tab1}. The feature extractors were separately-trained for OCT images of each acquisition factor $(l,m)$. Among the separately-trained models, the RCAN method reports the best PSNR and SSIM scores comparing to other methods for most of acquisition factors. Additionally, the MSRN method reports comparatively high PSNR and SSIM scores against RDN and RDU, whereas the number of trainable parameters in MSRN is significantly less than other methods. The RDU model does not excel in our OCT dataset even if it has more trainable parameters than RCAN. One possible explanation is that variations of U-Net have bottleneck layers, which might restrain the information flow in reconstruction tasks for OCT images\cite{9433801}.

In Table \ref{table:tab1}, we also report performance of meta-learning integrated with the best feature extractor RCAN (referred as MSSMN). \textcolor{black}{To make a fair comparison, we consistently set the magnification factors $(\hat{l}, \hat{m})$ equal to acquistion factors $(l,m)$ for MSSMN.} The MSSMN achieves the highest PSNR and SSIM scores among all acquisition factors. Using the same feature extractor, meta-learning brings an average improvement of 0.0674 in PSNR and an average improvement of 0.0015 in SSIM towards all acquisition factors from RCAN to MSSMN. For larger acquisition factors (less data acquired), the improvements are more significant. Overall, the combination of RCAN and meta-learning framework achieves the best performance for OCT image reconstruction in our dataset.

\textcolor{black}{We observe that the SSIM scores for reconstructed OCT images are lower compared to that of nature images \cite{Hu2019MetaSRAM,X.Hu.2020}. This is consistent with existing reports on OCT image reconstruction ([12], [13], [14], [16]) due to the comparatively low image quality. The high resolution OCT has lower image quality than high resolution natural images in terms of signal to noise ratio and image contrast. As the high resolution OCT image is with lower image quality than natural image as a reference to calculate SSIM and PSNR, we expect the SSIM and PSNR are lower in OCT-related tasks than those in natural image-related tasks. In addition, our acquisition strategy on both spatial and spectral domain brings extra challenges for MSSMN compared to downsampling OCT data in a single domain.}
\begin{table*}[h]
 \caption{PSNR And SSIM Results Of Reconstruction From Oct Images Of Acquisition Factor $(l,m)$ Using RCAN$(\hat{l},\hat{m})$. All Results Are Averaged Based On Five-Fold Cross-Validation. \textcolor{red}{Red} and \textcolor{blue}{\uline{Blue}} Indicate The Best And The Second Best Performance, Respectively.
 }
 \label{table:tab4}
 \centering
 \begin{tabular}{lll|l|l|l|l|l|l|l|l} 
 \hline
 \diagbox{Method}{$(l,m)$} & Metrics                                                              & \multicolumn{1}{c}{(2,2)}                                                                                    & \multicolumn{1}{c}{(2,3)}                                                                                   & \multicolumn{1}{c}{(2,4)}                                                                                    & \multicolumn{1}{c}{(3,2)}                                                                  & \multicolumn{1}{c}{(3,3)}                                                                  & \multicolumn{1}{c}{(3,4)}                                                                  & \multicolumn{1}{c}{(4,2)}                                                                           & \multicolumn{1}{c}{(4,3)}                                                                                    & \multicolumn{1}{c}{(4,4)}                                                           \\ 
 \hline
 RCAN ($\hat{l} = 2,\hat{m} = 2$) & \begin{tabular}[c]{@{}l@{}}PSNR\\SSIM\end{tabular} & \begin{tabular}[c]{@{}l@{}}\textcolor{blue}{\uline{29.1395}}\\\textcolor{blue}{\uline{0.7570 }}\end{tabular} & \begin{tabular}[c]{@{}l@{}}26.7070\\0.5586\end{tabular}                                                      & \begin{tabular}[c]{@{}l@{}}25.9195\\0.4510\end{tabular}                                                      & \begin{tabular}[c]{@{}l@{}}26.0177\\0.6018\end{tabular}                                                      & \begin{tabular}[c]{@{}l@{}}25.2471\\0.4829\end{tabular}                                                      & \begin{tabular}[c]{@{}l@{}}24.8725\\0.4100\end{tabular}                                                      & \begin{tabular}[c]{@{}l@{}}24.1563\\0.4858\end{tabular}                                            & \begin{tabular}[c]{@{}l@{}}24.2241\\0.4274\end{tabular}                                                      & \begin{tabular}[c]{@{}l@{}}24.1444\\0.3794\end{tabular}                                             \\
 RCAN ($\hat{l} = 2,\hat{m} = 3$) & \begin{tabular}[c]{@{}l@{}}PNSR\\SSIM\end{tabular} & \begin{tabular}[c]{@{}l@{}}28.2387\\0.6848\end{tabular}                                                      & \begin{tabular}[c]{@{}l@{}}\textcolor{blue}{\uline{27.3534}}\\\textcolor{blue}{\uline{0.6024 }}\end{tabular} & \begin{tabular}[c]{@{}l@{}}26.1379\\0.4639\end{tabular}                                                      & \begin{tabular}[c]{@{}l@{}}25.4245\\0.5404\end{tabular}                                                      & \begin{tabular}[c]{@{}l@{}}24.5441\\0.4210\end{tabular}                                                      & \begin{tabular}[c]{@{}l@{}}25.1339\\0.4203\end{tabular}                                                      & \begin{tabular}[c]{@{}l@{}}23.5747\\0.4388\end{tabular}                                            & \begin{tabular}[c]{@{}l@{}}23.6788\\0.3771\end{tabular}                                                      & \begin{tabular}[c]{@{}l@{}}24.3677\\0.3874\end{tabular}                                             \\
 RCAN ($\hat{l} = 2,\hat{m} = 4$) & \begin{tabular}[c]{@{}l@{}}PNSR\\SSIM\end{tabular} & \begin{tabular}[c]{@{}l@{}}28.1176\\0.6579\end{tabular}                                                      & \begin{tabular}[c]{@{}l@{}}25.7341\\0.4659\end{tabular}                                                      & \begin{tabular}[c]{@{}l@{}}\uline{\textcolor{blue}{26.4421}}\\\uline{\textcolor{blue}{0.4876 }}\end{tabular} & \begin{tabular}[c]{@{}l@{}}25.1130\\0.5202\end{tabular}                                                      & \begin{tabular}[c]{@{}l@{}}24.5077\\0.4099\end{tabular}                                                      & \begin{tabular}[c]{@{}l@{}}25.2193\\0.4319\end{tabular}                                                      & \begin{tabular}[c]{@{}l@{}}23.4057\\0.4294\end{tabular}                                            & \begin{tabular}[c]{@{}l@{}}23.5806\\0.3692\end{tabular}                                                      & \begin{tabular}[c]{@{}l@{}}24.3706\\0.3952\end{tabular}                                             \\
 RCAN ($\hat{l} = 3,\hat{m} = 2$) & \begin{tabular}[c]{@{}l@{}}PNSR\\SSIM\end{tabular} & \begin{tabular}[c]{@{}l@{}}28.3228\\0.7006\end{tabular}                                                      & \begin{tabular}[c]{@{}l@{}}26.2502\\0.5066\end{tabular}                                                      & \begin{tabular}[c]{@{}l@{}}25.3176\\0.4033\end{tabular}                                                      & \begin{tabular}[c]{@{}l@{}}\textcolor{blue}{\uline{26.9176}}\\\textcolor{blue}{\uline{0.6023 }}\end{tabular} & \begin{tabular}[c]{@{}l@{}}25.7484\\0.4670\end{tabular}                                                      & \begin{tabular}[c]{@{}l@{}}25.1228\\0.3879\end{tabular}                                                      & \begin{tabular}[c]{@{}l@{}}25.4813\\\textcolor{red}{0.5065 }\end{tabular}                          & \begin{tabular}[c]{@{}l@{}}25.2145\\0.4334\end{tabular}                                                      & \begin{tabular}[c]{@{}l@{}}24.8207\\0.3740\end{tabular}                                             \\
RCAN ($\hat{l} = 3,\hat{m} = 3$) & \begin{tabular}[c]{@{}l@{}}PNSR\\SSIM\end{tabular} & \begin{tabular}[c]{@{}l@{}}27.9758\\0.6534\end{tabular}                                                      & \begin{tabular}[c]{@{}l@{}}25.7518\\0.4610\end{tabular}                                                      & \begin{tabular}[c]{@{}l@{}}25.8236\\0.4345\end{tabular}                                                      & \begin{tabular}[c]{@{}l@{}}26.2365\\0.5472\end{tabular}                                                      & \begin{tabular}[c]{@{}l@{}}\uline{\textcolor{blue}{26.2668}}\\\uline{\textcolor{blue}{0.5082 }}\end{tabular} & \begin{tabular}[c]{@{}l@{}}25.4906\\0.4108\end{tabular}                                                      & \begin{tabular}[c]{@{}l@{}}24.5388\\0.4591\end{tabular}                                            & \begin{tabular}[c]{@{}l@{}}24.6297\\0.3931\end{tabular}                                                      & \begin{tabular}[c]{@{}l@{}}25.0996\\0.3903\end{tabular}                                             \\
 RCAN ($\hat{l} = 3,\hat{m} = 4$) & \begin{tabular}[c]{@{}l@{}}PNSR\\SSIM\end{tabular} & \begin{tabular}[c]{@{}l@{}}27.6449\\0.6235\end{tabular}                                                      & \begin{tabular}[c]{@{}l@{}}25.8364\\0.4552\end{tabular}                                                      & \begin{tabular}[c]{@{}l@{}}26.0717\\0.4522\end{tabular}                                                      & \begin{tabular}[c]{@{}l@{}}25.5899\\0.5239\end{tabular}                                                      & \begin{tabular}[c]{@{}l@{}}25.1103\\0.4189\end{tabular}                                                      & \begin{tabular}[c]{@{}l@{}}\textcolor{blue}{\uline{25.7503}}\\\textcolor{blue}{\uline{0.4326 }}\end{tabular} & \begin{tabular}[c]{@{}l@{}}23.8753\\0.4464\end{tabular}                                            & \begin{tabular}[c]{@{}l@{}}24.3177\\0.3864\end{tabular}                                                      & \begin{tabular}[c]{@{}l@{}}25.0535\\\textcolor{red}{0.3999}\end{tabular}                            \\
 RCAN ($\hat{l} = 4,\hat{m} = 2$) & \begin{tabular}[c]{@{}l@{}}PNSR\\SSIM\end{tabular} & \begin{tabular}[c]{@{}l@{}}26.3729\\0.5671\end{tabular}                                                      & \begin{tabular}[c]{@{}l@{}}24.9097\\0.4074\end{tabular}                                                      & \begin{tabular}[c]{@{}l@{}}24.2386\\0.3392\end{tabular}                                                      & \begin{tabular}[c]{@{}l@{}}26.3620\\0.5373\end{tabular}                                                      & \begin{tabular}[c]{@{}l@{}}25.0821\\0.4005\end{tabular}                                                      & \begin{tabular}[c]{@{}l@{}}24.4809\\0.3390\end{tabular}                                                      & \begin{tabular}[c]{@{}l@{}}\uline{\textcolor{blue}{25.9547}}\\0.4973\end{tabular}                  & \begin{tabular}[c]{@{}l@{}}25.0280\\0.3922\end{tabular}                                                      & \begin{tabular}[c]{@{}l@{}}24.5449\\0.3400\end{tabular}                                             \\
 RCAN ($\hat{l} = 4,\hat{m} = 3$) & \begin{tabular}[c]{@{}l@{}}PNSR\\SSIM\end{tabular} & \begin{tabular}[c]{@{}l@{}}27.4361\\0.6086\end{tabular}                                                      & \begin{tabular}[c]{@{}l@{}}25.6006\\0.4377\end{tabular}                                                      & \begin{tabular}[c]{@{}l@{}}25.4481\\0.4052\end{tabular}                                                      & \begin{tabular}[c]{@{}l@{}}26.4774\\0.5318\end{tabular}                                                      & \begin{tabular}[c]{@{}l@{}}25.3258\\0.4132\end{tabular}                                                      & \begin{tabular}[c]{@{}l@{}}25.3649\\0.3913\end{tabular}                                                      & \begin{tabular}[c]{@{}l@{}}25.3811\\0.4617\end{tabular}                                            & \begin{tabular}[c]{@{}l@{}}\textcolor{blue}{\uline{25.7287}}\\\textcolor{blue}{\uline{0.4486 }}\end{tabular} & \begin{tabular}[c]{@{}l@{}}25.1836\\0.3793\end{tabular}                                             \\
 RCAN ($\hat{l} = 4,\hat{m} = 4$) & \begin{tabular}[c]{@{}l@{}}PNSR\\SSIM\end{tabular} & \begin{tabular}[c]{@{}l@{}}27.3637\\0.5875\end{tabular}                                                      & \begin{tabular}[c]{@{}l@{}}25.7250\\0.4350\end{tabular}                                                      & \begin{tabular}[c]{@{}l@{}}25.7403\\0.4216\end{tabular}                                                      & \begin{tabular}[c]{@{}l@{}}26.1471\\0.5135\end{tabular}                                                      & \begin{tabular}[c]{@{}l@{}}25.4230\\0.4120\end{tabular}                                                      & \begin{tabular}[c]{@{}l@{}}25.6337\\0.4054\end{tabular}                                                      & \begin{tabular}[c]{@{}l@{}}24.6886\\0.4489\end{tabular}                                            & \begin{tabular}[c]{@{}l@{}}25.0076\\0.3901\end{tabular}                                                      & \begin{tabular}[c]{@{}l@{}}\textcolor{blue}{\uline{25.4509}}\\0.3947\end{tabular}                   \\
 MSSMN ($\hat{l} = l,  \hat{m} = m$) & \begin{tabular}[c]{@{}l@{}}PNSR\\SSIM\end{tabular} & \begin{tabular}[c]{@{}l@{}}\textcolor{red}{29.1789}\\\textcolor{red}{0.7578}\end{tabular} & \begin{tabular}[c]{@{}l@{}}\textcolor{red}{27.3914}\\\textcolor{red}{0.6040}\end{tabular} & \begin{tabular}[c]{@{}l@{}}\textcolor{red}{26.4562}\\\textcolor{red}{0.4876}\end{tabular} & \begin{tabular}[c]{@{}l@{}}\textcolor{red}{27.0109}\\\textcolor{red}{0.6031}\end{tabular} & \begin{tabular}[c]{@{}l@{}}\textcolor{red}{26.3555}\\\textcolor{red}{0.5104}\end{tabular} & \begin{tabular}[c]{@{}l@{}}\textcolor{red}{25.8655}\\\textcolor{red}{0.4333}\end{tabular} & \begin{tabular}[c]{@{}l@{}}\textcolor{red}{26.0577}\\\textcolor{blue}{\uline{0.4993}}\end{tabular} & \begin{tabular}[c]{@{}l@{}}\textcolor{red}{25.7946}\\\textcolor{red}{}\textcolor{red}{0.4507}\end{tabular} & \begin{tabular}[c]{@{}l@{}}\textcolor{red}{25.4995}\\\uline{\textcolor{blue}{0.3978}}\end{tabular} \\
 \hline
 \end{tabular}
 \end{table*}

\subsection{Performance comparison using a unified network on different acquisition factors}
We tested the flexibility of using a unified network to reconstruct images when magnification factors $(\hat{l},\hat{m})$ were different from acquisition factors $(l,m)$. We compared the performance of using MSSMN with other RCAN models that were trained for single magnification factors. The RCAN$(\hat{l},\hat{m})$ stands for a RCAN model that is trained for magnification factor of $(\hat{l},\hat{m})$. The output images of RCAN$(\hat{l}, \hat{m})$ were resized to the reference OCT images in order to calculate the PSNR and SSIM scores. \textcolor{black}{The magnification factors $(\hat{l}, \hat{m})$ for MSSMN were equal to acquisition factors $(l,m)$.}

As shown in Table \ref{table:tab4}, the MSSMN reports higher PSNR and SSIM scores for resolving OCT images of most acquisition factors. 
We find that the best performance of RCAN$(\hat{l}, \hat{m})$ generally occurs when ideally $l=\hat{l}$ and $m=\hat{m}$. Also, the second best PSNR and SSIM scores for a RCAN$(\hat{l},\hat{m})$ model are more likely to occur where at least one of the acquisition factors (either in spatial or spectral domain) matches $(\hat{l},\hat{m})$. 
For acquisition factor $(4,4)$, among RCAN$(\hat{l},\hat{m})$ models, the best PSNR (25.4509) is achieved by RCAN$(4,4)$, and the best SSIM (0.5095) is achieved by RCAN$(4,3)$.
For acquisition factor $(4,2)$, among RCAN$(\hat{l},\hat{m})$ models, the best PSNR (25.9547) is achieved by RCAN$(4,2)$, and the best SSIM (0.5065) is achieved by RCAN$(3,2)$. Overall, Table \ref{table:tab4} also supports our argument that the output of RCAN$(\hat{l},\hat{m})$ models will be deteriorated if applied to OCT images of different acquisition factors $(l,m)$. In contrast,
training a unified network, MSSMN, generally achieves better performance than training each RCAN-based model to handle various magnification factors 
\textcolor{black}{due to the nature of multi-task learning. Reconstructing OCT using multiple magnification factors can be considered as solving multiple tasks. Useful information is exploited from related tasks during multi-task learning, which in turn improves learning procedure for each task \cite{9392366}. The meta-upscaling and meta-restoration modules in MSSMN provide platforms for multi-task learning, which leads to the improvements in each magnification factor compared to baseline RCAN models trained for specific acquisition factors.}

\subsection{Performance comparison on OCT images of untrained acquisition factors}
\begin{table*}[t]
\caption{Results Of Reconstructed Oct Images By Magnification Factor $(\lfloor\hat{l}\rfloor, \lfloor \hat{m} \rfloor)$ Using RCAN and Magnification Factor $(\hat{l}, \hat{m})$ Using MSSMN. All Results Are Averaged Based On Five-Fold Cross-Validation. \textcolor{red}{Red} and \textcolor{blue}{\uline{Blue}} Indicate The Best And The Second Best Performance, Respectively.}
\label{table:tab5}
\centering
\setlength\tabcolsep{5pt}
\begin{tabular}{lll|l|l|l|l|l|l|l|l}
\hline
\diagbox{Method}{($\hat{l},\hat{m}$)} & \multicolumn{1}{l}{Metrics} & \multicolumn{1}{c}{(2.5,2.5)} & \multicolumn{1}{c}{(2.5,3.5)} & \multicolumn{1}{c}{(2.5,4.5)} & \multicolumn{1}{c}{(3.5,2.5)} & \multicolumn{1}{c}{(3.5,3.5)} & \multicolumn{1}{c}{(3.5,4.5)} & \multicolumn{1}{c}{(4.5,2.5)} & \multicolumn{1}{c}{(4.5,3.5)} & \multicolumn{1}{c}{(4.5,4.5)} \\
\hline
RCAN & \begin{tabular}[c]{@{}l@{}}PSNR\\SSIM\end{tabular} & \begin{tabular}[c]{@{}l@{}}\uline{\textcolor{blue}{26.3234}}\\\uline{\textcolor{blue}{0.5760}}\end{tabular} & \begin{tabular}[c]{@{}l@{}}\uline{\textcolor{blue}{25.4072}}\\\uline{\textcolor{blue}{0.4441}}\end{tabular} & \begin{tabular}[c]{@{}l@{}}\uline{\textcolor{blue}{25.2823}}\\\uline{\textcolor{blue}{0.4011}}\end{tabular} & \begin{tabular}[c]{@{}l@{}}\uline{\textcolor{blue}{25.6820}}\\\uline{\textcolor{blue}{0.4899}}\end{tabular} & \begin{tabular}[c]{@{}l@{}}\uline{\textcolor{blue}{25.1737}}\\\uline{\textcolor{blue}{0.4034}}\end{tabular} & \begin{tabular}[c]{@{}l@{}}\textcolor{blue}{\uline{25.0225}}\\\textcolor{blue}{\uline{0.3735}}\end{tabular} & \begin{tabular}[c]{@{}l@{}}\textcolor{blue}{\uline{25.3419}}\\\textcolor{blue}{\uline{0.4284}}\end{tabular} & \begin{tabular}[c]{@{}l@{}}\textcolor{blue}{\uline{25.0736}}\\\textcolor{blue}{\uline{0.3797}}\end{tabular} & \begin{tabular}[c]{@{}l@{}}\textcolor{blue}{\uline{25.0012}}\\\textcolor{blue}{\uline{0.3559}}\end{tabular} \\
MSSMN & \begin{tabular}[c]{@{}l@{}}PSNR\\SSIM\end{tabular} & \begin{tabular}[c]{@{}l@{}}\textcolor{red}{27.2603}\\\textcolor{red}{0.6123 }\end{tabular} & \begin{tabular}[c]{@{}l@{}}\textcolor{red}{26.3926}\\\textcolor{red}{0.5005 }\end{tabular} & \begin{tabular}[c]{@{}l@{}}\textcolor{red}{25.8637}\\\textcolor{red}{0.4236 }\end{tabular} & \begin{tabular}[c]{@{}l@{}}\textcolor{red}{26.1884}\\\textcolor{red}{0.5103 }\end{tabular} & \begin{tabular}[c]{@{}l@{}}\textcolor{red}{25.7945}\\\textcolor{red}{0.4398 }\end{tabular} & \begin{tabular}[c]{@{}l@{}}\textcolor{red}{25.4855}\\\textcolor{red}{0.3887 }\end{tabular} & \begin{tabular}[c]{@{}l@{}}\textcolor{red}{25.6679}\\\textcolor{red}{0.4469 }\end{tabular} & \begin{tabular}[c]{@{}l@{}}\textcolor{red}{25.4557}\\\textcolor{red}{0.4047 }\end{tabular} & \begin{tabular}[c]{@{}l@{}}\textcolor{red}{25.2380}\\\textcolor{red}{0.3677 }\end{tabular} \\
\hline
\end{tabular}
\end{table*}

The choice of acquisition factors $(l,m)$ should be adjustable based on needs in coronary analysis. However, it is technically impossible to train the reconstruction framework for OCT images of all possible acquisition factors, especially the non-integer factors. Thus, it is desirable to propose a framework that is robust and able to generalize for untrained magnification factors, especially the factors which are non-integers. 

We used the MSSMN to reconstruct OCT images of untrained acquisition factors $(l,m)$, here $l=2.5, 3.5, 4.5$ and $m=2.5, 3.5, 4.5$. \textcolor{black}{The floor operation is used to achieve non-integer spatial acquisition factor $m$. Using floor operation, the A-lines indexes are adjusted while the desired 
spatial acquisition factor $m$ can be achieved.} To compare the performances, we also calculated the PSNR and SSIM scores of reconstruction using RCAN$(\lfloor l \rfloor, \lfloor m \rfloor)$. For example, we used the RCAN model trained for magnification factor $(2,2)$ to resolve OCT images of acquisition factor $(2.5,2.5)$ if $l=2.5$ and $m=2.5$. 
With the meta-upscaling and meta-restoration modules, our MSSMN is capable of magnifying OCT images of untrained acquisition factors. In Table \ref{table:tab5}, MSSMN method reports higher PSNR and SSIM scores compared to the baseline RCAN$(\lfloor l \rfloor, \lfloor m \rfloor)$ models. The differences of scores between RCAN and MSSMN are larger in resolving OCT images of untrained acquisition factors compared to that of trained factors (shown in Table \ref{table:tab1}), which means our framework is generalizable.

\subsection{Visualization of drug-eluting stents in coronary imaging}
\begin{figure*}[t]
\centering
\includegraphics[width=18cm]{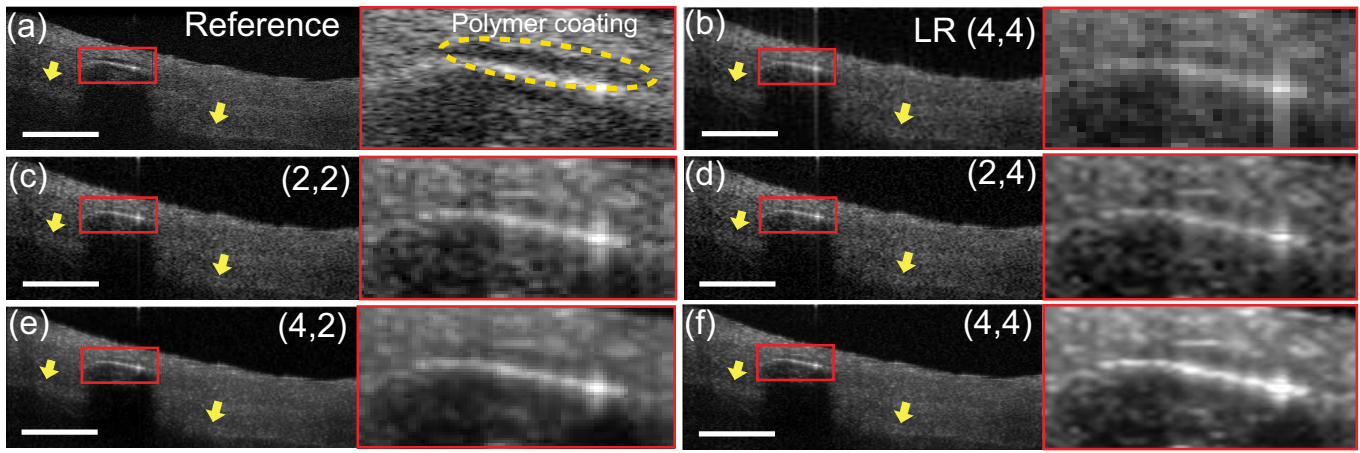}
\caption{Multi-scale image reconstruction of drug-eluting stents. (a) The reference OCT image. (b) The LR OCT image is down-scaled with an acquisition factor of $(4,4)$. (c)-(f) The reconstructed images using different magnification factors $(2,2)$, $(2,4)$, $(4,2)$ and $(4,4)$. Scale bar: 200 $\mu$m. Stent alloy, polymer coating and boundary between media and adventitia can be resolved using different magnification factors. }
\label{fig:multiScaleHistology}
\end{figure*}

\begin{figure*}[h]
\centering
\includegraphics[width=18cm]{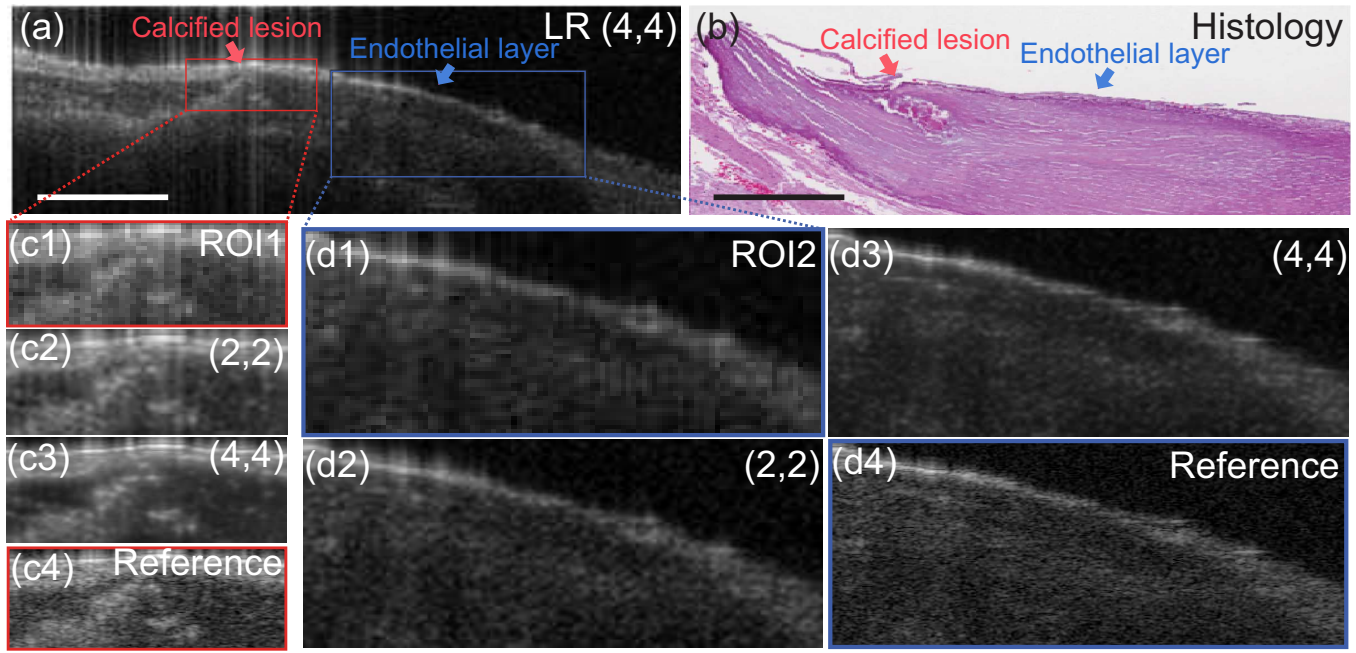}
\caption{Multi-scale image reconstruction of two ROIs. (a) The LR OCT image is down-scaled with an acquisition factor of $(4,4)$. (b) Histology image, (c1-c3,d1-d3) ROIs. (c4,d4) The reference OCT images. Scale bar: 200 $\mu$m. ROIs are marked by red and blue rectangles respectively. ROIs are magnified by MSSMN using magnification factors of $(2,2)$ and $(4,4)$. For ROI1, a magnification factor of $(2,2)$ resolves general shape of calcified lesion; For ROI2, a magnification factor of $(4,4)$ provides better assessment of the thickness of endothelial layer.}
\label{fig:picture004}
\end{figure*}
Imaging microstructures and tissues adjacent to stent struts is crucial in clinic. It is critical provides accurate morphological information of interactions between the stent and vessel wall. With drug-eluting stents (DESs), polymer coatings elute drugs, which prevents intimal proliferation as well as reduces the risk of restenosis after stent implantation. Polymer coating on DESs usually appears as clear rims of material around the metallic struts\cite{LinboLiu.2011}. After image down-scaling, such information might be compromised in LR OCT images.
\begin{figure*}[h]
\centering
\includegraphics[width=18cm]{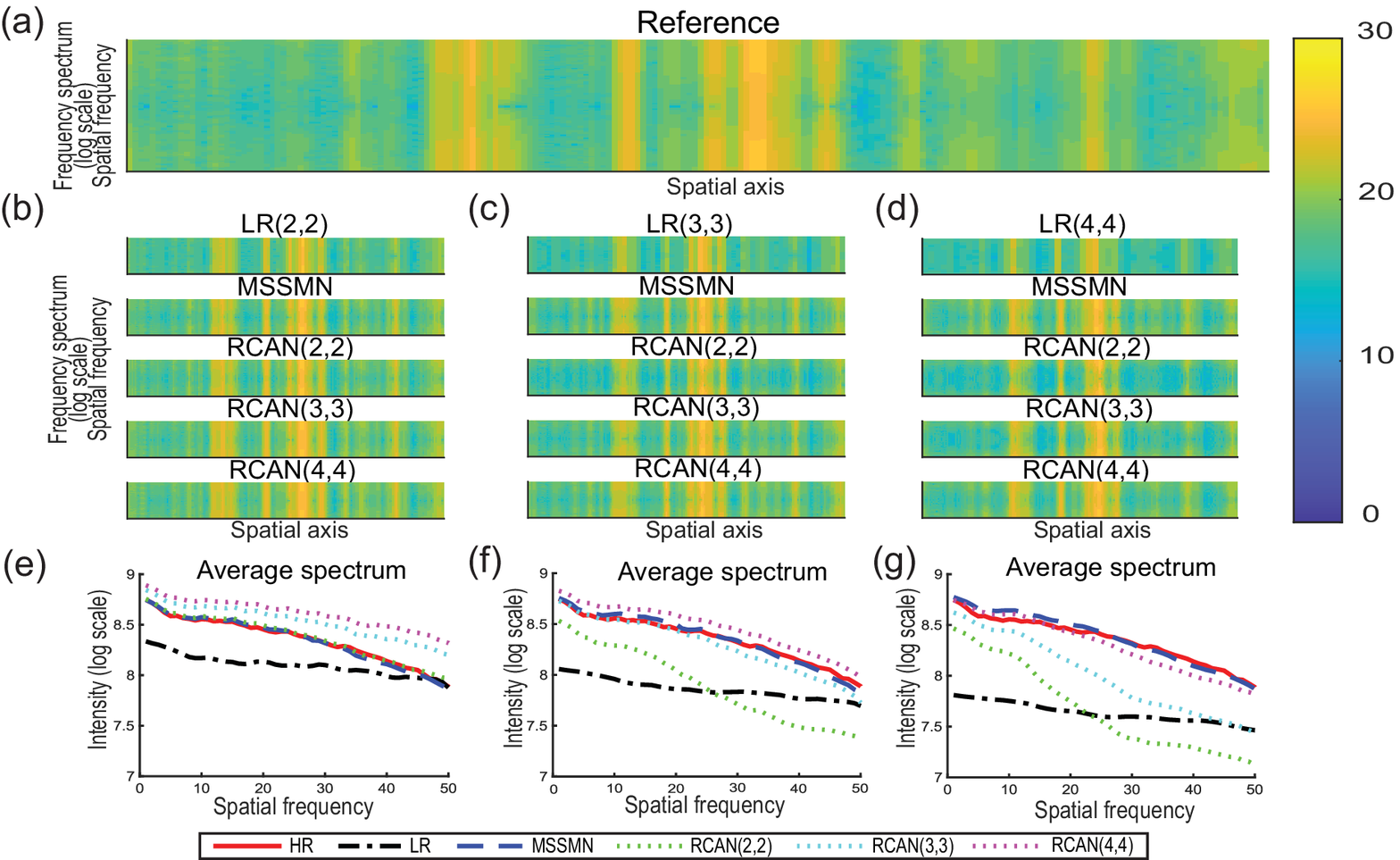}
\caption{Spatial frequency analysis of the reference, LR and reconstructed OCT images. (a) Log-scaled spatial frequency spectra of the reference zoomed ROI1 in Fig. \ref{fig:picture004}. (b)-(d) First row: Log-scaled spatial frequency spectra of OCT images compressed by factors $(2,2)$, $(3,3)$ and $(4,4)$. Second row: outcome from MSSMN. Third row to Fifth row: outcome from RCAN$(2,2)$, RCAN$(3,3)$, and RCAN$(4,4)$. (e)-(g) Averaged intensity of spectral profiles in (b)-(d). Our framework, MSSMN, generates OCT images that better match the spectrum of the reference image.}
\label{fig:picture005}
\end{figure*}

In Fig. \ref{fig:multiScaleHistology}(a), we demonstrate the appearance of polymer coating on a stent captured by OCT. However, the polymer coating and stent alloy are hardly visible in compressed OCT image with a acquisition factor of $(4,4)$ as shown in Fig. \ref{fig:multiScaleHistology}(b). We applied our framework to the compressed OCT image, magnifying them by different factors in both spectral and spatial domains. With a magnification factor of $(2,2)$, the structure of stent alloy was restored as shown in Fig. \ref{fig:multiScaleHistology}(c). However, polymer coating on the stent was still blurry. Thus, we used a higher magnification factor in spatial domain in order to mitigate the blurring artifacts, as shown in Fig. \ref{fig:multiScaleHistology}(d). With the magnification factor of $(2,4)$, blurry artifacts were removed thus the polymer coating can be better visualized. In this case, a magnification factor $(2,4)$ should be enough if we are interested in the stent and polymer coating. But for more detailed textures such as boundary between media and adventitia (highlighted by yellow arrows), a higher magnification factor in spatial domain is required. In Fig. \ref{fig:multiScaleHistology}(e), we used a magnification factor $(4,2)$ to resolve the compressed OCT image, which provided information about stent, polymer coating, and the boundary between media and adventitia. Further with a magnification factor of $(4,4)$, the detailed textures were better recovered with a higher magnification factor in terms of digital resolution as shown in Fig. \ref{fig:multiScaleHistology}(f).

In Fig. \ref{fig:multiScaleHistology}, we demonstrated performance of our framework using multiple magnification factors to resolve a compressed OCT image containing important features including stent, polymer coating, and boundary between media and adventitia. The result suggests that resolving compressed OCT images with multiple magnification factors in both spectral and spatial domain provides flexibility which suits structures with various sizes.

\subsection{Visual inspection and assessment with H\&E stained histological micrographs}

To demonstrate flexibility of our framework in magnifying pathological features, we applied our framework to two magnified regions of interests (ROIs), ROI1 and ROI2, in an example of low resolution (LR) OCT image acquired using a factor of $(4,4)$. 

We applied magnification factors of $(2,2)$ and $(4,4)$ to magnify two ROIs. The visual inspections are shown in Fig. \ref{fig:picture004}. In ROI1, a calcified lesion is highlighted. For ROI1, a magnification factor $(2,2)$ is clear enough to draw the general shape of calcified region. In ROI2, the thickness of endothelial layer needs to be precisely assessed. For ROI2, a magnification factor $(4,4)$ is necessary for an accurate assessment of the thickness of endothelium. In conventional setup, two networks need to be separately trained to magnify two scales. On the contrary, our framework can freely switch between magnification factors using a unified network. We thus provide flexibility for analyzing pathological features within coronary OCT images.

\subsection{Comparison of output spectrum}
To further quantify our network inference results against the ground truth images, we performed spatial frequency analysis to the reference OCT image, LR OCT images, reconstruction results from MSSMN and RCAN models. The reference image is a zoomed ROI, the same on as shown in the red box of Fig. \ref{fig:picture004}(a). We magnified each LR image using MSSMN, RCAN$(2,2)$, RCAN$(3,3)$ and RCAN$(4,4)$. The magnification factor of MSSMN is equal to the acquisition factor of LR.

The LR OCT images were acquired using acquisition factors of $(2,2)$, $(3,3)$ and $(4,4)$. We converted the images into the spatial frequency domain using 1D Fourier transform for each A-line (along the vertical axis). Results of spatial frequency analysis are shown in Fig. \ref{fig:picture005}. The results of spatial frequency comparison for each A-line, as shown in Fig. \ref{fig:picture005}(b)-(d), demonstrate that our framework is capable of providing OCT images that closely match the reference in the spectral domain. The detailed quantitative comparison of spectrum are given in Fig. \ref{fig:picture005}(e)-(g). Moreover, the separately trained RCAN models with fixed magnification factor $(\hat{l},\hat{m})$, if applied to compressed OCT image of different acquisition factors, outcomes diverged spectrum. The spatial frequency analysis supports our argument that the separately-trained RCAN models are less robust and can hardly be applied to compressed OCT images of different factors.

\section{Discussion}
We propose a new image acquisition method and a reconstruction framework to achieve fast image visualization at flexible magnification factors. We devise a new image acquisition method in spectral-spatial domain for SD-OCT systems, which improves acquisition speed and reduces data storage consumption while maintaining high image quality in reconstruction. To reconstruct the compressed OCT images of different acquisition factors, we develop a multi-scale framework which functions in both spectral and spatial domains. Our framework, MSSMN, achieves better PSNR and SSIM scores comparing to that of separately-trained MSRN, RDN, RDU and RCAN models. Experimental results suggest that separately-trained RCAN models report deteriorated PSNR and SSIM scores if applied to OCT images of different acquisition factors. We confirm that the proposed multi-scale reconstruction framework is beneficial for supporting our acquisition method. Moreover, our framework achieves higher PSNR and SSIM scores when adopting untrained magnification factors comparing to that of RCAN models. Our viable acquisition method and multi-scale reconstruction framework are particularly valuable to the development of catheter based IVOCT systems. The image acquisition speed of SD-OCT systems can be improved without any hardware modifications using our acquisition method, which makes it more suitable to be applied in intravascular applications. Meanwhile, the resolving power of SD-OCT systems is maintained by adopting our multi-scale reconstruction framework. We have demonstrated the flexibility of our multi-scale framework in resolving pathological features. 

\textcolor{black}{We compare our work with existing OCT reconstruction works from three aspects: data acquisition strategy, neuron network, and data size. We present a unique spectral-spatial OCT data acquisition method which modifies OCT images in both spectral and spatial domains. According to our experiment, our acquisition method outperforms uniform sampling when the spectral acquisition $l$ factor is larger than 2. Moreover, we have performed extensive reconstruction experiments on four feature extractors for comparison: MSRN, RDN, RDU, and RCAN. We demonstrated RCAN is the optimal selection of feature extractor for MSSMN. Lastly, our research is carried out on a reasonably large dataset. In terms of deep learning-based OCT super-resolution or enhancement, the training data size used in our work is larger than that of [14] (100 images), [16] (500 images), [20] (320 images), [21] (1388 images), [22] (128 images), [23] (25 images), [25] (280 images) and [26] (429 images). }

There are a few directions to further extend our framework in the future. First, our data acquisition method and multi-scale reconstruction network can be easily adopted in SS-OCT systems. SS-OCT and SD-OCT share similar principle in acquiring spectral data and use Fourier Transform for image reconstruction. Second, the image reconstruction framework can be integrated in a GAN structure. There is a potential to further improve image quality by adding a discriminator in current neural network. \textcolor{black}{Third, we plan to incorporate de-blurring and de-noising modules in MSSMN with special considerations on blurring and noise corruption.} Finally, we will extend the study from benchtop OCT data acquisition to catheter-based OCT data acquisition.

\section{Conclusion}
In this paper, we propose a spectral-spatial based sampling method to boost efficiency of image acquisition while maintaining high performance and flexibility in reconstruction. We develop a multi-scale reconstruction framework, MSSMN, for compressed OCT images. Our experimental results suggest that the framework outperforms separately-trained networks. 
We demonstrate that our framework is capable of reconstructing OCT images and magnifying pathological structures with flexible magnification factors as determined by clinical needs. With extensive experiments in human coronary OCT images, we demonstrate high accuracy, data efficiency, and magnification flexibility. Our work may allow for faster SD-OCT inspection with high resolution during coronary intervention.

\section*{Acknowledgment}
The authors would like to thank Kelsey Nagle for data acquisition, Dr. Fei Hu for technical discussion, and Dr. Dezhi Wang for histology service.

\bibliographystyle{ieeetr}
\bibliography{reference}

\begin{thebibliography}{10}

\bibitem{benjamin2019heart}
E.~J. Benjamin {\em et~al.}, ``Heart disease and stroke statistics—2019
  update: a report from the american heart association,'' {\em Circulation},
  vol.~139, no.~10, pp.~e56--e528, 2019.

\bibitem{benjo2015high}
A.~M. Benjo {\em et~al.}, ``High dose statin loading prior to percutaneous
  coronary intervention decreases cardiovascular events: a meta-analysis of
  randomized controlled trials,'' {\em Catheterization and Cardiovascular
  Interventions}, vol.~85, no.~1, pp.~53--60, 2015.

\bibitem{wijns2015optical}
W.~Wijns {\em et~al.}, ``Optical coherence tomography imaging during
  percutaneous coronary intervention impacts physician decision-making:
  {ILUMIEN I} study,'' {\em European heart journal}, vol.~36, no.~47,
  pp.~3346--3355, 2015.

\bibitem{meneveau2016optical}
N.~Meneveau {\em et~al.}, ``Optical coherence tomography to optimize results of
  percutaneous coronary intervention in patients with non--st-elevation acute
  coronary syndrome: results of the multicenter, randomized doctors study (does
  optical coherence tomography optimize results of stenting),'' {\em
  Circulation}, vol.~134, no.~13, pp.~906--917, 2016.

\bibitem{jones2018angiography}
D.~A. Jones {\em et~al.}, ``Angiography alone versus angiography plus optical
  coherence tomography to guide percutaneous coronary intervention: outcomes
  from the pan-london pci cohort,'' {\em JACC: Cardiovascular Interventions},
  vol.~11, no.~14, pp.~1313--1321, 2018.

\bibitem{tearney2008imaging}
G.~J. Tearney, I.-K. Jang, and B.~Bouma, ``Imaging coronary atherosclerosis and
  vulnerable plaques with optical coherence tomography,'' in {\em Optical
  Coherence Tomography}, pp.~1083--1101, Springer, 2008.

\bibitem{LinboLiu.2011}
L.~Liu {\em et~al.}, ``Imaging the subcellular structure of human coronary
  atherosclerosis using micro--optical coherence tomography,'' {\em Nature
  Medicine}, vol.~17, no.~8, pp.~1010--1014, 2011.

\bibitem{Fang.2013}
L.~Fang {\em et~al.}, ``Fast acquisition and reconstruction of optical
  coherence tomography images via sparse representation,'' {\em IEEE
  Transactions on Medical Imaging}, vol.~32, no.~11, pp.~2034--2049, 2013.

\bibitem{Fang.2017}
L.~Fang {\em et~al.}, ``Segmentation based sparse reconstruction of optical
  coherence tomography images,'' {\em IEEE Transactions on Medical Imaging},
  vol.~36, no.~2, pp.~407--421, 2017.

\bibitem{Abbasi.2018}
A.~Abbasi {\em et~al.}, ``Optical coherence tomography retinal image
  reconstruction via nonlocal weighted sparse representation,'' {\em Journal of
  biomedical optics}, vol.~23, no.~3, pp.~1--11, 2018.

\bibitem{Daneshmand.2021}
P.~G. Daneshmand, A.~Mehridehnavi, and H.~Rabbani, ``Reconstruction of optical
  coherence tomography images using mixed low rank approximation and second
  order tensor based total variation method,'' {\em IEEE Transactions on
  Medical Imaging}, vol.~40, no.~3, pp.~865--878, 2021.

\bibitem{Zhang.2021b}
Y.~Zhang {\em et~al.}, ``Neural network-based image reconstruction in
  swept-source optical coherence tomography using undersampled spectral data,''
  {\em Light: Science {\&} Applications}, vol.~10, no.~1, p.~155, 2021.

\bibitem{KaichengLiang.2020}
K.~Liang {\em et~al.}, ``Resolution enhancement and realistic speckle recovery
  with generative adversarial modeling of micro-optical coherence tomography,''
  {\em Biomed. Opt. Express}, vol.~11, no.~12, pp.~7236--7252, 2020.

\bibitem{Z.Yuan.2020}
{Z. Yuan} {\em et~al.}, ``Axial super-resolution study for optical coherence
  tomography images via deep learning,'' {\em IEEE Access}, vol.~8,
  pp.~204941--204950, 2020.

\bibitem{Tearney.2012}
G.~J. Tearney {\em et~al.}, ``Consensus standards for acquisition, measurement,
  and reporting of intravascular optical coherence tomography studies: a report
  from the international working group for intravascular optical coherence
  tomography standardization and validation,'' {\em Journal of the American
  College of Cardiology}, vol.~59, no.~12, pp.~1058--1072, 2012.

\bibitem{AntoniaLichtenegger.2021}
A.~Lichtenegger {\em et~al.}, ``Reconstruction of visible light optical
  coherence tomography images retrieved from discontinuous spectral data using
  a conditional generative adversarial network,'' {\em Biomed. Opt. Express},
  vol.~12, no.~11, pp.~6780--6795, 2021.

\bibitem{DBLP:journals/corr/SimonyanZ14a}
K.~Simonyan and A.~Zisserman, ``Very deep convolutional networks for
  large-scale image recognition,'' in {\em 3rd International Conference on
  Learning Representations, {ICLR} 2015, San Diego, CA, USA, May 7-9, 2015,
  Conference Track Proceedings} (Y.~Bengio and Y.~LeCun, eds.), 2015.

\bibitem{Ronneberger.2015}
O.~Ronneberger, P.~Fischer, and T.~Brox, ``U-net: Convolutional networks for
  biomedical image segmentation,'' in {\em Medical Image Computing and
  Computer-Assisted Intervention -- MICCAI 2015} (N.~Navab, J.~Hornegger, W.~M.
  Wells, and A.~F. Frangi, eds.), (Cham), pp.~234--241, {Springer International
  Publishing}, 2015.

\bibitem{PengfeiGuo.2020}
P.~Guo, D.~Li, and X.~Li, ``Deep {OCT} image compression with convolutional
  neural networks,'' {\em Biomed. Opt. Express}, vol.~11, no.~7,
  pp.~3543--3554, 2020.

\bibitem{Xu.2018}
Y.~Xu {\em et~al.}, ``Improving the resolution of retinal {OCT} with deep
  learning,'' in {\em Medical Image Understanding and Analysis} (M.~Nixon,
  S.~Mahmoodi, and R.~Zwiggelaar, eds.), (Cham), pp.~325--332, {Springer
  International Publishing}, 2018.

\bibitem{Hao.2020b}
Q.~Hao {\em et~al.}, ``High signal-to-noise ratio reconstruction of low
  bit-depth optical coherence tomography using deep learning,'' {\em Journal of
  biomedical optics}, vol.~25, no.~12, p.~123702, 2020.

\bibitem{Qiu2020N2NSROCTSD}
B.~Qiu {\em et~al.}, ``N2nsr-oct: Simultaneous denoising and super-resolution
  in optical coherence tomography images using semi-supervised deep
  learning.,'' {\em Journal of biophotonics}, p.~e202000282, 2020.

\bibitem{HongmingPan.2020b}
H.~Pan {\em et~al.}, ``More realistic low-resolution {OCT} image generation
  approach for training deep neural networks,'' {\em OSA Continuum}, vol.~3,
  no.~11, pp.~3197--3205, 2020.

\bibitem{S.Cao.2020}
{S. Cao} {\em et~al.}, ``Super-resolution technology to simultaneously improve
  optical {\&} digital resolution of opticalsuper-resolution technology to
  simultaneously improve optical {\&} digital resolution of opticalhe coherence
  tomography via deep learning,'' in {\em 2020 42nd Annual International
  Conference of the IEEE Engineering in Medicine {\&} Biology Society (EMBC)},
  pp.~1879--1882, 2020.

\bibitem{Zhou.2020}
T.~Zhou {\em et~al.}, ``Digital resolution enhancement in low transverse
  sampling optical coherence tomography angiography using deep learning,'' {\em
  OSA Continuum}, vol.~3, no.~6, p.~1664, 2020.

\bibitem{Das.2020}
V.~Das, S.~Dandapat, and P.~Bora, ``Unsupervised super-resolution of oct images
  using generative adversarial network for improved age-related macular
  degeneration diagnosis,'' {\em IEEE Sensors Journal}, vol.~PP, p.~1, 2020.

\bibitem{W.Yang.2019}
{W. Yang} {\em et~al.}, ``Deep learning for single image super-resolution: A
  brief review,'' {\em IEEE Transactions on Multimedia}, vol.~21, no.~12,
  pp.~3106--3121, 2019.

\bibitem{YulunZhang.2018b}
Y.~Zhang {\em et~al.}, ``Image super-resolution using very deep residual
  channel attention networks,'' in {\em ECCV}, 2018.

\bibitem{Wang.2019b}
X.~Wang {\em et~al.}, ``{ESRGAN}: Enhanced super-resolution generative
  adversarial networks,'' in {\em Computer Vision -- ECCV 2018 Workshops}
  (L.~Leal-Taix{\'e} and S.~Roth, eds.), (Cham), pp.~63--79, {Springer
  International Publishing}, 2019.

\bibitem{Hu2019MetaSRAM}
X.~Hu and othersn, ``{Meta-SR}: A magnification-arbitrary network for
  super-resolution,'' {\em 2019 IEEE/CVF Conference on Computer Vision and
  Pattern Recognition (CVPR)}, pp.~1575--1584, 2019.

\bibitem{X.Hu.2020}
{X. Hu} {\em et~al.}, ``{Meta-USR}: A unified super-resolution network for
  multiple degradation parameters,'' {\em IEEE Transactions on Neural Networks
  and Learning Systems}, pp.~1--15, 2020.

\bibitem{7797130}
H.~Zhao {\em et~al.}, ``Loss functions for image restoration with neural
  networks,'' {\em IEEE Transactions on Computational Imaging}, vol.~3, no.~1,
  pp.~47--57, 2017.

\bibitem{Li.2018c}
J.~Li {\em et~al.}, ``Multi-scale residual network for image
  super-resolution,'' in {\em Computer Vision -- ECCV 2018} (V.~Ferrari,
  M.~Hebert, C.~Sminchisescu, and Y.~Weiss, eds.), (Cham), pp.~527--542,
  {Springer International Publishing}, 2018.

\bibitem{Zhang2018ResidualDN}
Y.~Zhang {\em et~al.}, ``Residual dense network for image super-resolution,''
  {\em 2018 IEEE/CVF Conference on Computer Vision and Pattern Recognition},
  pp.~2472--2481, 2018.

\bibitem{GurrolaRamos.2021}
J.~Gurrola-Ramos, O.~Dalmau, and T.~E. Alarc{\'o}n, ``A residual dense u-net
  neural network for image denoising,'' {\em IEEE Access}, vol.~9,
  pp.~31742--31754, 2021.

\bibitem{Shi2016RealTimeSI}
W.~Shi {\em et~al.}, ``Real-time single image and video super-resolution using
  an efficient sub-pixel convolutional neural network,'' {\em 2016 IEEE
  Conference on Computer Vision and Pattern Recognition (CVPR)},
  pp.~1874--1883, 2016.

\bibitem{kingma2014adam}
D.~P. Kingma and J.~Ba, ``Adam: A method for stochastic optimization,'' {\em
  arXiv preprint arXiv:1412.6980}, 2014.

\bibitem{ZhouWang.2004b}
Z.~Wang {\em et~al.}, ``Image quality assessment: from error visibility to
  structural similarity,'' {\em IEEE Transactions on Image Processing},
  vol.~13, no.~4, pp.~600--612, 2004.

\bibitem{9433801}
S.~Lee and I.~V. Bajić, ``{Information Flow Through U-Nets},'' in {\em 2021
  IEEE 18th International Symposium on Biomedical Imaging (ISBI)},
  pp.~812--816, 2021.

\bibitem{9392366}
Y.~Zhang and Q.~Yang, ``A survey on multi-task learning,'' {\em IEEE
  Transactions on Knowledge and Data Engineering}, pp.~1--1, 2021.

\end{thebibliography}

%






\end{document}